# Spin glass behavior in frustrated quantum spin system $CuAl_2O_4$ with a possible orbital liquid state

R. Nirmala[1,2,3], Kwang-Hyun Jang[1,3], Hasung Sim[4,5], Hwanbeom Cho[4,5], Junghwan Lee[1,3], Nam-Geun Yang[1,3], Seongsu Lee[6,7], R. M. Ibberson[8,9], K. Kakurai[10], M. Matsuda[10,11], S.-W. Cheong[7], V.V. Gapontsev[12] and S.V. Streltsov[12,13] and Je-Geun Park[1,3,4,5]∗

[1]*Center for Strongly Correlated Materials Research, Seoul National University, Seoul 08826, Korea*
[2]*Department of Physics, Indian Institute of Technology Madras, Chennai 600 036, India*
[3]*Department of Physics, Sungkyunkwan University, Suwon 16419 Korea*
[4]*Center for Correlated Electron Systems, Institute for Basic Science, Seoul 08826, Korea*
[5]*Department of Physics and Astronomy, Seoul National University, Seoul 08826, Korea*
[6]*Neutron Science Division, Korea Atomic Energy Research Institute, Daejeon 34057, Korea*
[7]*Rutgers Center for Emergent Materials and Department of Physics and Astronomy, Rutgers University, Piscataway, New Jersey 08854, USA*
[8]*ISIS Facility, Rutherford Appleton Laboratory, Didcot OX11 0QX, UK*
[9]*Chemical and Engineering Materials, Oak Ridge National Laboratory, Oak Ridge, Tennessee 37831, USA*
[10]*Quantum Beam Science Center, Japan Atomic Energy Agency, Tokai, Ibaraki 319-1195, Japan*
[11]*Quantum Condensed Matter Division, Oak Ridge National Laboratory, Oak Ridge, Tennessee 37831, USA*
[12]*M.N. Miheev Institute of Metal Physics of Ural Branch of Russian Academy of Sciences, 620137, Ekaterinburg, Russia*
[13]*Ural Federal University, Mira St. 19, 620002 Ekaterinburg, Russia*

∗ Corresponding author
E-mail: jgpark10@snu.ac.kr

**Abstract**

$CuAl_2O_4$ is a normal spinel oxide having quantum spin, S=1/2 for $Cu^{2+}$. It is a rather unique feature that the $Cu^{2+}$ ions of $CuAl_2O_4$ sit at a tetrahedral position, not like the usual octahedral position for many oxides. At low temperatures, it exhibits all the thermodynamic evidence of a quantum spin glass. For example, the polycrystalline $CuAl_2O_4$ shows a cusp centered at ~2 K in the low-field dc magnetization data and a clear frequency dependence in the ac magnetic susceptibility while it displays logarithmic relaxation behavior in a time dependence of the magnetization. At the same time, there is a peak at ~2.3 K in the heat capacity, which shifts towards higher temperature with magnetic fields. On the other hand, there is no evidence of new superlattice peaks in the high-resolution neutron powder diffraction data when cooled from 40 to 0.4 K. This implies that there is no long-ranged magnetic order down to 0.4 K, thus confirming a spin glass-like ground state for $CuAl_2O_4$. Interestingly, there is no sign of structural distortion either although $Cu^{2+}$ is a Jahn-Teller active ion. Thus, we claim that an orbital liquid state is the most likely ground state in $CuAl_2O_4$. Of further interest, it also exhibits a large frustration parameter, $f = |\theta_{CW}/T_m| \sim 67$, one of the largest values reported for spinel oxides. Our observations suggest that $CuAl_2O_4$ should be a rare example of a frustrated quantum spin glass with a good candidate for an orbital liquid state.

**Introduction**

A magnetic system with bond or site disorder often exhibits a short-ranged order, indicating that the system cannot form a true thermodynamic ground state and thus becomes frustrated. This state of matter, so-called spin glass, with a multitude of a ground state degeneracy has drawn considerable interest over the past few decades [1,2]. More recently, there have been intensive studies on a quantum version of the spin glass, in particular, in a lattice with $Cu^{2+}$ ions, and a more exotic quantum spin liquid phase.

The $AB_2O_4$ (where A = divalent cation such as Cu, Fe and Co; B = trivalent cation such as Al and Fe) family of spinel oxides is known for their unique magnetic properties. In fact, their magnetic properties highly depend upon the cation distribution in the basic cubic structure (space group *Fd-3m*, no. 227). Often, an inversion parameter ($x$) defined as the amount of trivalent cations at the tetrahedral sites (*8a*) is used to identify whether a spinel oxide forms in one of either 'largely normal' or 'largely inverse' structures [3]. It is usually accepted among the community that a 'normal spinel' structure is formed for $0 < x < 2/3$ while an 'inverse spinel' structure is stabilized with $2/3 < x < 1$. While the A-site of the spinel structure forms a diamond lattice, the B-site forms a pyrochlore lattice. This pyrochlore lattice hosts a three-dimensional network of corner-sharing tetrahedra, which for antiferromagnetic interactions gives rise to intrinsic geometric frustration leading to novel emergent properties such as spin ice, cluster glass, etc. [4,5,6]. Some of spinel compounds with magnetic ions at the B-site show glassy behavior too [7,8].

On the other hand, normal spinel oxides with magnetic ions at the A-site are expected to have a collinear antiferromagnetic ground state [9]. In these oxides, A-A nearest-neighbors (NN) interaction is stronger as compared with A-B and B-B interactions.

In fact, being bipartite a diamond structure does not, a priori, have frustration with nearest-neighbor interaction alone, regardless whether it be antiferromagnetic or ferromagnetic. Nevertheless, strong magnetic frustration effects were observed in the physical properties of a few normal sulphospinels such as $MnSc_2S_4$ and $FeSc_2S_4$ [10]. This then led to an idea that these systems could be explained by a Hamiltonian including next-nearest-neighbor (NNN) antiferromagnetic exchange interaction [11]. Theoretical studies have also shown that with this NNN interaction the normal A-site spinel can exhibit various degenerate magnetic ground states [11]. It should be noted that some A-site normal spinels such as $FeAl_2O_4$ and $CoAl_2O_4$ are known to display glass-like behavior due to the competition between the NN and NNN interactions for the diamond structure of the A-sublattice [11,12,13].

Several experimental studies have so far been made to achieve a better understanding of the ground state magnetic properties of $TAl_2O_4$ (T = Co, Fe and Mn) oxides with the normal spinel structure. For example, while $CoAl_2O_4$ and $FeAl_2O_4$ display spin glass-like ground states, $FeAl_2O_4$ shows a sign of an additional orbital freezing at lower temperatures. On the other hand, $MnAl_2O_4$ shows a long-ranged antiferromagnetic order below 40 K [13,14,15]. Among these $TAl_2O_4$ systems, $CoAl_2O_4$ is the most frustrated material lying close to a quantum critical point of an antiferromagnetic state [12,16].

By comparison, there have been relatively very few studies on the magnetic properties of copper-based normal spinel oxides and their physical properties remain largely unexplored. Naturally, the quantum spin (S=1/2) of $Cu^{2+}$ makes it an attractive candidate, which might harbor an exotic quantum magnetic ground state. Therefore, it will be interesting to examine the ground state of the Cu-based S=1/2 systems in detail. Of several Cu-based spinel compounds, $CuGa_2O_4$ is reported to have an inverse spinel structure and to exhibit a spin-glass state below 2.5 K [17, 18]. A brief report [18] was also made about the non-zero values of a zero-point entropy for both $CuGa_2O_4$ and $CuAl_2O_4$, indicative of unusual ground state properties. We also note that nanostructured $CuAl_2O_4$ has been studied for possible photocatalytic applications [19]. However, to our best knowledge there is no report of the physical properties of $CuAl_2O_4$.

This work focuses on the Cu-based A-site spinel oxide of $CuAl_2O_4$ to find glassy behavior with an extremely large frustration parameter of *f*=67. Interestingly, the frustration seems to result from the competing NN and NNN antiferromagnetic interactions supported by the underlying diamond-like structure. The spin glass order might as well be driven by small cation disorders that are difficult to remove completely in the normal spinel oxides. Of particularly interest is the fact that our high-resolution neutron diffraction data do not show any sign of structural distortion down to 0.4 K although $Cu^{2+}$ is a Jahn-Teller active ion. This implies that $CuAl_2O_4$ might host a possibly quantum spin glass state in the background of an orbital liquid ground state.

**Experimental Details**

The polycrystalline $CuAl_2O_4$ sample was prepared by a solid state reaction method, starting from pure CuO and $Al_2O_3$ (99.99% pure, Sigma-Aldrich) [20]. The samples were heated at 900° C for 9 hours with intermediate grindings and then sintered at 1020° C for 24 hours in air. We also prepared $ZnAl_2O_4$ under similar conditions to use it as a non-magnetic system for the analysis of our heat capacity data. We initially checked the quality of the sample by X-ray diffraction experiments (Miniflex II, Rigaku). There were small (less than 1%) impurity of $Al_2O_3$. DC magnetization measurements were performed using a SQUID magnetometer (MPMS, Quantum Design, USA) in fields up to 5 T in the temperature range of 1.8 – 300 K. We also used a $He^3$ insert of the SQUID magnetometer to measure dc magnetization data down to 0.5 K. In order to study further the frequency dependence, we measured ac susceptibility under an ac field of 0.3 mT using the SQUID magnetometer at different frequencies from 1 to 1,000 Hz in the temperature range of 1.8 – 150 K with both zero and small dc bias fields. Magnetic relaxation measurements were carried out using the SQUID magnetometer under a standard protocol: we collected the data under 5 T as a function of time after zero-field cooling (ZFC).

Heat capacity was measured by a relaxation technique from 0.5 to 300 K with a commercial system (PPMS, Quantum Design, USA). Powder neutron diffraction (ND) experiments were carried out down to 0.4 K using two neutron instruments: one is the high-resolution time-of-flight diffractometer (HRPD) at the ISIS Facility, UK and the other is a cold triple-axis spectrometer, TAS2 of JAEA, Japan operated in a two-axis mode between 1.45 and 10 K. We also made separate high-resolution X-ray diffraction measurement using a commercial machine (D8 advance, Bruker) from 300 to 40 K equipped with a low temperature cryostat: FullProf was used for structure analysis [21].

**Results and Discussion**

Our high-resolution neutron and X-ray powder diffraction data confirm that our $CuAl_2O_4$ sample forms in a 'largely normal' spinel structure (see Figs. 1a & 4a) with $Cu^{2+}$ ions at the A-site and $Al^{3+}$ ions at the B-site (space group *Fd-3m*, No. 227). See Table 1 for the summary of the structural information. Our estimate of the lattice parameter of 8.07683 (5) Å is close to those values reported in the previous works: 8.078(1) and 8.079 Å [22, 23]: we note that Ref 18 reported a smaller value of the lattice parameter, 8.045 Å. Our full structure analysis was carried out by the joint refinement of both neutron HRPD data and HR-XRD data taken at 40 and 300 K. The final refinement data show a significant site-inversion for $Cu^{2+}$ ions, which occupy the otherwise forbidden octahedral B-sites (see Table 1). We note that a smaller (6 ~ 8 %) amount of site inversion was reported for other polycrystalline normal spinel oxides $TAl_2O_4$ (T = Co, Fe and Mn) [13,14]. It is to be noted too that a similar amount (8 %) of the site inversion was recently reported for $CoAl_2O_4$ single crystals [16]. All these experimental results including ours indicate how much difficult it is to remove the small site inversion completely for these $TAl_2O_4$ systems.

In order to study the low temperature magnetic properties, we measured the dc magnetization of $CuAl_2O_4$ with an applied field of 5 mT down to 0.5 K (see the inset of Fig. 1b). As one can see, there is a clear cusp centered at 2 K ($T_m$), indicative of a magnetic phase transition. When measured under both field-cooled and zero-field-cooled conditions, the data show a clear bifurcation behavior. This is a typical sign of spin glass behavior. As we will show later, the position of this peak moves toward higher temperature with increasing frequency. Thus, the irreversible behavior in the dc magnetization can be taken as evidence of the spin glass transition occurring at low temperature for $CuAl_2O_4$. At higher temperatures, the dc magnetization follows the Curie-Weiss law with the Curie-Weiss temperature of $\theta_{CW}$ ~ -137 K and the effective moment value of ~1.95 $\mu_B/Cu^{2+}$ (Fig. 1b).

This estimate of the effective moment value compares very well with the theoretical spin-only value of $Cu^{2+}$ ion (1.73 $\mu_B$). We comment that the measured $\theta_{CW}$ is extremely large compared to the cusp temperature ($T_m$) with $f = 67$, where the frustration parameter ($f$) is defined as $f = |\theta_{CW}/T_m|$. When the frustration parameter is larger than 5 ~ 10, it is generally considered as a highly frustrated magnet [11]. Thus, it is striking that we find $f$ as large as 67 for $CuAl_2O_4$. It was previously reported that Fe-doped $CuAl_2O_4$ exhibits signs of a spin glass state [24].

To put the large $f$ value for $CuAl_2O_4$ in perspective, it is useful to compare it with those for other Al spinel compounds: $f$ ~10 - 22 for $CoAl_2O_4$ (S = 3/2), $f$ ~11 for $FeAl_2O_4$ (S = 2) and $f$ ~3.6 for $MnAl_2O_4$ (S= 5/2) [13,14]. For comparison, $f$ is found to be only ~3 for another Cu spinel compound $CuGa_2O_4$ although it has the same quantum spin of $Cu^{2+}$ like $CuAl_2O_4$ [17]. Therefore, it is expected that over the wide temperature range of $T_m \leq T \leq |\theta_{CW}|$ $CuAl_2O_4$ fluctuates between many low-energy configurations, evading a long-ranged order. It is also interesting to note that the $f$ parameter appears to fall with increasing spin **S** value and with increasing exchange interactions for $TAl_2O_4$ (T = Co, Fe and Mn) oxides.

Our heat capacity data measured under 0 and 9 Tesla also confirm the weak magnetic order. The C/T vs T plot shows a broad peak centered at ~2.3 K in zero field, consistent with the observation of a weak magnetic transition in the susceptibility (Fig. 2). This peak shifts towards higher temperature with applied magnetic fields and reaches at ~ 4 K with 9 T. This shift of the peak position with increasing magnetic field is also consistent with a spin glass nature as we will discuss shortly. In order to model the experimental observation of the low-temperature peak, we used a Schottky model (line) in Fig. 2a with a gap value of $\Delta$=6.5 K: $C = Nk_B \left(\frac{\Delta}{k_B T}\right)^2 \frac{e^{\Delta/k_B T}}{\left(1+e^{\Delta/k_B T}\right)^2}$. As one can see, the Schottky model describes the experimental data reasonably well.

Upon close inspection, the heat capacity data show a nonlinear temperature dependence below $T_m$ with a power of ~ 2 unlike that observed in canonical spin glass systems [2]. However, we note that $T^2$ dependence was also found in CoAl$_2$O$_4$ [13]. A similar $T^2$ dependence has also been observed in other geometrically frustrated spin glass with a Kagome structure [25]. Interestingly, a fractional power law ($T^{2.33}$) dependence was theoretically predicted for a low-temperature heat capacity in a glassy state based on classical treatments of a diamond lattice Heisenberg antiferromagnet after including competing NN and NNN interactions [11].

For further quantitative analysis of the data, we obtained the magnetic part of the heat capacity, $C_m$, by subtracting the heat capacity of the non-magnetic analogue ZnAl$_2$O$_4$ from the experimental data of CuAl$_2$O$_4$ (Fig. 2). From the $C_m$/T vs T data, we estimated a magnetic entropy as a function of temperature (see Fig. 2b). According to our estimate the magnetic entropy only reaches S=0.843 J/mol K even at 20 K, which is far smaller than the theoretical maximum molar magnetic entropy value of Cu$^{2+}$ ion, Rln2 = 5.763 J/mol K. This difference leads us to a value of the zero-point entropy of 4.92 J/mol K, which is very close to the value reported in Ref. [18]. If we consider the orbital degeneracy of Cu$^{2+}$ ion with t$_{2g}$ manifold, this theoretical value should even get increased by the additional orbital entropy of Rln3. The application of magnetic field shifts most of the magnetic entropy from low to high temperatures as expected (Fig. 2). It is interesting to note that the magnetic entropy of $FeCr_2S_4$ single crystal, a candidate for an orbital glass state, reaches just over 2 J/mol-K at 20 K, which is again much smaller than the theoretical value of the combined spin and orbital entropy, Rln2+Rln5 [26].

The deficit of the magnetic entropy as compared with the theoretical value can be in principle interpreted in two ways. First, it implies that some significant parts of the total entropy may be released at much lower temperature or resided at zero temperature. Another equally possible scenario is that the total entropy may be recovered at much higher temperatures although we should note that we already integrated the entropy to 20 K, which is 10 times higher than the actual transition temperature. Either way, it implies strongly that the low temperature phase of $CuAl_2O_4$ is nontrivial.

To further examine the nature of the weak magnetic transition, we measured the ac susceptibility of $CuAl_2O_4$ under an ac field of 0.3 mT at different frequencies ranging from 1 Hz to 1 kHz. The real part of the ac susceptibility (Fig. 3a) demonstrates a clear frequency dependence below 2.5 K: the peak seen at 2 K for 1 Hz shifts towards higher temperatures with increasing frequency and eventually reaches at ~2.4 K for 1 kHz. This shift of the peak temperature with increasing frequency is considered as a hallmark of spin glass, as often observed in materials with magnetic frustration or competing interactions [1, 2, 27].

The magnetization vs field data of $CuAl_2O_4$ show a linear field dependence at 50 and 300 K (Fig. 3b). However, a S-shaped curvature develops at lower temperatures (inset in Fig. 3b). From the data measured at 0.6 K under the magnetic field of 5 T, we estimated the magnetic moment value to be ~ 0.12 $\mu_B/Cu^{2+}$ at 5 T but the magnetization never reaches a saturation. We also note that there is a small but finite non-zero isothermal remnant magnetization (IRM) at 0.6 K. The three observations: the S-shaped magnetization, no saturation behavior, and the IRM, are consistent with the typical behavior of canonical spin glass [2] and were previous noted for other oxides too [27].

While the frequency-dependent ac susceptibility confirms the spin glass state and its nonequilibrium characteristics, we further looked for a possible slow relaxation of the magnetization at low temperatures and performed magnetic relaxation measurements on $CuAl_2O_4$. For this we specifically measured the magnetization as a function of time (t), M(t), and plotted the data in the inset of Fig. 3a after normalizing the data with respect to the initial magnetization at t = 0, i.e. M(t)/M(0). For this relaxation measurement, the sample was initially zero-field-cooled to each target temperature from 300 K. After a magnetic field of 5 T was switched on and the field was stabilized (t = 0), we started to collect the time-dependent data. As one can see in the figure, the magnetic relaxation effects are clearly observable below 10 K; which becomes most pronounced at 2 K, coinciding with the cusp temperature in the dc and ac magnetization data (Figs. 1 & 3). The magnetization still shows relaxation behavior even after waiting for more than an hour at 2 K. Note that time was measured in seconds in the inset of Fig. 3a. The magnetic relaxation behavior is a canonical feature of spin glass [1,2,28]. We further comment that the magnetization ratio is almost linear in the log(t) plots in the glassy region: which is also a characteristics of a conventional spin glass, indicating the evolution of a broad distribution of spin relaxation rates in the vicinity of the freezing temperature [29].

Although all the bulk data discussed above consistently show that the low-temperature transition is of spin glass origin, one needs further microscopic studies such as neutron diffraction experiment. In order to investigate the temperature evolution of the crystal as well as magnetic structures, we carried out high-resolution neutron powder diffraction experiments from 300 to 0.4 K using the HRPD beamline of the ISIS facility, UK. Our data collected at 40 K can be well fitted with the normal spinel structure as discussed above (see Table 1). Most interestingly, a direct comparison of the data with that taken at 0.4 K does not show any new superlattice peaks at the low temperatures (see the difference curve of Fig. 4b). This is clear evidence in favor of the absence of any long-ranged order in $CuAl_2O_4$. We further note that a separate measurement using a triple-axis-spectrometer, TAS2 of JAEA, also produced a very similar picture, i.e. no sign of magnetic ordering in the low-temperature diffraction data, reinforcing our view that the magnetic transition seen in $CuAl_2O_4$ is not of a long-ranged order.

Of further interest is also the fact that our data collected both above and below the transition temperature failed to show any sign of structural distortion: all our diffraction data are consistent with the cubic Fd-3m space group. This is quite surprising given the fact that $Cu^{2+}$ is a Jahn-Teller active ion. To further demonstrate this point, we show the enlarged picture of the (4 0 0) and (2 2 2) Bragg peaks as the insets in Fig. 4b: both peaks have more or less the same value of FWHM (Full Width at Half-Maximum) at 40 and 0.4 K: for the (400) peak it is FWHM=0.00353(4) at 0.4 K and FWHM= 0.00349(4) at 40 K; for the (222) peak it is FWHM=0.00371(3) at 0.4 K and FWHM=0.00375(3) at 40 K. Furthermore, we carried out high-resolution XRD experiments using a commercial machine with the resolutions of 0.04° (Bruker D8 Discover).

As a separate analysis of possible structural distortions, we simulated the diffraction patterns for the (400) and (222) peaks with a hypothetical tetragonal distortion of 0.5% elongation along the c-axis. As one can see in the dashed-line next to the data in the inset of Fig. 4b, this rather small tetragonal distortion split the (400) peaks, which is easily detectable by our high-resolution neutron diffraction experiment. Thus we can safely put the upper bound of a possible Jahn-Teller distortion for $CuAl_2O_4$ at lower than 0.1%. However, it is still possible that $Cu^{2+}$ ions undergo a local Jahn-Teller distortion without breaking the global symmetry. Local probes such as a total scattering technique might be helpful to answer this question although our high-resolution diffraction data seem to put it in doubt.

Before moving on to the theoretical discussion as to the absence of the Jahn-Teller distortion in $CuAl_2O_4$, we would like to point out a subtle, but importance difference for the tetrahedral symmetry as compared to the octahedral configuration. The usual Jahn-Teller distortion of ions at the octahedral surrounding, e.g. perovskite oxides, is defined as a distance between a metal (Me) ion and oxygen (O). However, the Jahn-Teller distortions for the tetrahedra symmetry is not measured by a Me-O distance, but by a O-Me-O angle. This difference between O-Me-O angles then translates to the c/a ratio, which is the hallmark of the Jahn-Teller distortions for ions at the tetrahedral configuration. This value is usually not that small and can be easily detectable by X-ray or neutron diffractometer with reasonable resolutions: e.g. the c/a ratio is 0.91 for both $CuCr_2O_4$ and $CuRh_2O_4$ and about 1.04 for $NiCr_2O_4$ and $NiRh_2O_4$ [30]. As the measure of local distortions, we calculate two O-Me-O angles for $CuRh_2O_4$, 122.6° and 103.3°, and the difference between the two angles, i.e. the strength of the local distortion for $CuRh_2O_4$, is as large as 19.3° while it is zero to the accuracy of better than 0.1% for $CuAl_2O_4$.

As the $Cu^{2+}$ ion experiences the tetrahedral crystal field of neighboring four oxygen ions, it is expected to have three $t_{2g}$ levels ($d_{xy}$, $d_{xz}$, and $d_{yz}$) separated from the two low-lying $e_g$ levels ($d_{x^2-y^2}$ and $d_{z^2}$). Therefore, there is one hole in the $t_{2g}$ manifold for $CuAl_2O_4$, which is susceptible to a further Jahn-Teller distortion to lower the total energy [31]. The absence of Jahn-Teller distortion in our sample strongly suggests that a possibly orbital liquid state is being realized for $CuAl_2O_4$. (See Fig. 1a for our schematic diagram for an orbital liquid-like state within the diamond structure of $CuAl_2O_4$). We note that for octahedral $Cu^{2+}$ the Jahn-Teller distortion yields a low-symmetry distortion around the transition-metal ion and increases the stabilization energy [32,33,34].

In order to check this idea, we performed *ab initio* calculations within the local density approximation (LDA) and the linearized muffin-tin orbitals (LMTO) method [35]. Using the Wannier function projection procedure [36], we estimated hopping parameters between different $t_{2g}$ orbitals. Since these orbitals are degenerate, there are no dominating elements in the hopping matrix. The Frobenius norm of the hopping matrix for nearest neighbor Cu ions $||t||_{nn}$ = 130 meV. This results in a strong exchange coupling $J_{nn} = 4t^2/U$ ~ 110 K ($U$~7 eV in cuprates [37]), which should give a very large Neel temperature in a strong contrast to experimental findings. The origin of a strong suppression of the ordering temperature in $CuAl_2O_4$ is in a strong frustration of the exchange interaction. Indeed, our LDA calculations show that next nearest neighbor Cu, $||t||_{nnn}$ = 110 meV, is only slightly smaller than for nearest neighbors $||t||_{nn}$ = 130 meV. Then, the ratio between corresponding exchange constants $J_2/J_1$, characterizing degree of frustration, is of order of *(//t//$_{nnn}$ ///t//$_{nn}$)$^2$*~0.7, which implies strong suppression of the ordering temperature [11] and additionally supports the possibility of the orbital liquid state in $CuAl_2O_4$. This theoretical calculation support the idea too that the combination of the NN and NNN interactions gives rise to the magnetic frustration of $CuAl_2O_4$.

It is of further interest to recall that $Ba_3CuSbO_9$ having a honeycomb lattice system with $Cu^{2+}$ (S=1/2) ions does not show any sign of Jahn-Teller distortion. It has thus been interpreted as another geometrically frustrated system with spin-orbital liquid properties [38,39]. We also note that as $Cu^{2+}$ with the $t_{2g}$ manifold is rather rare among Cu oxides $CuAl_2O_4$ offers a new opportunity of exploring quantum magnetism with the $t_{2g}$ physics.

A passing comment, $Cu^{2+}$ ions with partially filled $e_g$-orbitals experience the strongest Jahn-Teller distortions in octahedral surrounding. The main reason is that the $e_g$-orbitals are directed to ligands, oxygen in our case, whereas $t_{2g}$-orbitals look as much as possible away. As a result, a large difference of 0.2~0.3 Å exists between long and short bonds in compounds like $LaMnO_3$ and $KCuF_3$ while in $LaTiO_3$ and $YTiO_3$ having a single electron in $t_{2g}$ orbitals exhibit a distortion, an order of magnitude smaller [30]. For the tetrahedral symmetry, the situation is just opposite. As a result the coupling of electronic subsystem with the lattice is weaker for $Cu^{2+}$ in the tetrahedral configuration and the Jahn-Teller distortion are expected to be much smaller.

Last, not least, we would like to make a general comment. Unlike many Cu spin systems, including high-Tc Cu superconductors, our study on $CuAl_2O_4$ offers a rare case, in which Cu ions sit at a tetrahedral configuration. To our view, this difference in the local chemistry makes our studies rather unique. This tetrahedral symmetry with $t_{2g}$ manifolds having one hole exhibits much weaker Jahn-Teller distortion simply because of the generic wave function of $t_{2g}$ orbitals. We believe that this point is very important to realizing the orbital liquid phase.

**Summary**

To summarize, a large frustration parameter is observed in the A-site magnetic spinel oxide $CuAl_2O_4$, with $Cu^{2+}$ (S=1/2). All the bulk measurements such as frequency-dependent ac susceptibility, magnetic relaxation behavior and specific heat indicate a spin glass transition around 2 K, consistent with the neutron diffraction data. We speculate that this spin glass order might possibly be aided by the small cation disorder in addition to the competing NN and NNN interactions. Most interestingly, our high-resolution neutron powder diffraction experiments do not show any sign of structural transition. The lack of structural distortion despite $Cu^{2+}$ ion being Jahn-Teller active indicates that an orbital liquid state might be realized for $CuAl_2O_4$. Thus, the low temperature phase of $CuAl_2O_4$ can be considered as a candidate for a quantum spin glass on the orbital liquid lattice.

**Acknowledgements**

We acknowledge D. Khomskii and Y. B. Kim for useful discussion, and the Quantum Design, Japan, for allowing us to use their $He^3$ option of heat capacity set-up. R.N would like to thank the IBS-CCES for support during her summer visits. Work at IIT Madras was supported by the institute NFSC and ERP schemes while work at IBS-CCES was supported by IBS-R009-G1. The work at Rutgers University was supported by the DOE under Grant No. DOE: DE-FG02-07ER46382. The work at Ekaterinburg was supported by the RFBR via 16-32-60070 program and by the FASO via theme ``Electron'' No. 01201463326.

**Figure Captions**

Fig.1 (Color online) (a) The spinel crystal structure and a schematic diagram of $t_{2g}$ orbitals showing the random distribution of $Cu^{2+}$ $d$-states within the diamond structure of $CuAl_2O_4$. (b) Inverse susceptibility vs temperature of $CuAl_2O_4$ taken with an applied field of 0.1 T: the Curie-Weiss fit (line) with the Curie-Weiss temperature ($\theta_{CW}$) of -137 K and the effective moment of 1.95 $\mu_B/Cu^{2+}$. Inset in (b) depicts the magnetization data collected down to 0.6 K after both field-cooling and zero-field-cooling using a $He^3$ insert of the SQUID magnetometer.

Fig. 2 (Color online) (a) Plot of C/T vs T of $CuAl_2O_4$ measured under applied fields of 0 and 9 T together with data (bottom) taken for $ZnAl_2O_4$. The inset shows the $T^2$ dependence of low temperature heat capacity data with lines as guide for eyes. We used a Schottky model (line) in order to model the experimental observation of the low-temperature peak. See the text. (b) It shows the temperature dependence of the magnetic entropy.

Fig. 3 (Color online) (a) Real part of ac susceptibility ($\chi$) vs temperature at different frequencies taken with an applied field of 0.3 mT. The arrow indicates the direction of increasing frequency. The inset shows the relaxation behavior of magnetization ratio M(t)/M(0) as a function of time taken at a few selected temperatures under applied field of 5 T. (b) Magnetization vs field data of $CuAl_2O_4$ at a few representative temperatures: 300, 50, 2, and 0.6 K, in applied fields up to 5 T with the inset showing the low field region of the 0.6 K data.

Fig. 4 (Color online) (a) Neutron powder diffraction data of $CuAl_2O_4$ taken at 40 K. The symbols represent the data points with the line displaying the refinement results. The vertical ticks indicate the position of expected Bragg peaks with the difference curve shown at the bottom. (b) The difference curve between the neutron diffraction data obtained at 0.4 and 40 K shows no sign of any new superlattice peaks of magnetic origin. The insets in Fig. 4b are the enlarged pictures of the (4 0 0) and (2 2 2) Bragg peaks with similar values of FWHM, demonstrating that there is no structural distortion. We also simulated the diffraction patterns for (400) and (222) peaks for a hypothetical tetragonal distortion of 0.5% elongation along the c-axis (see the line in the inset).

Figure 1

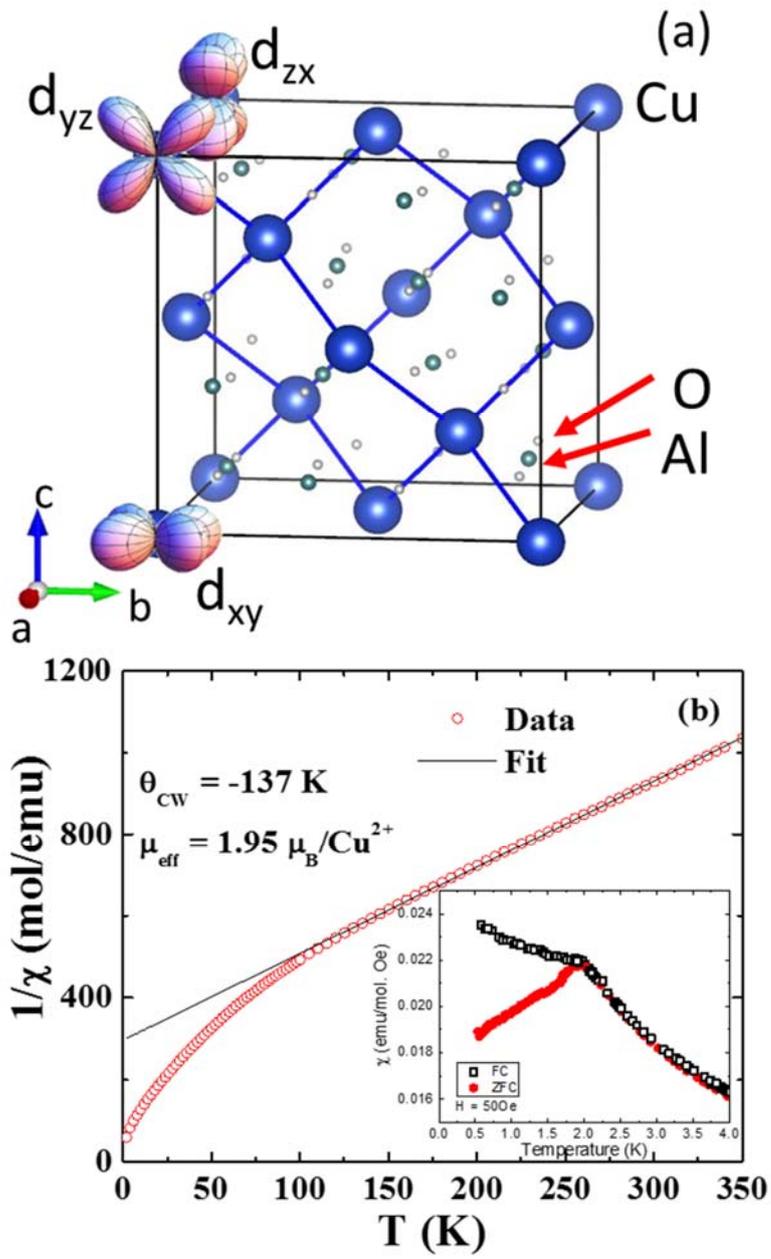

Figure 2

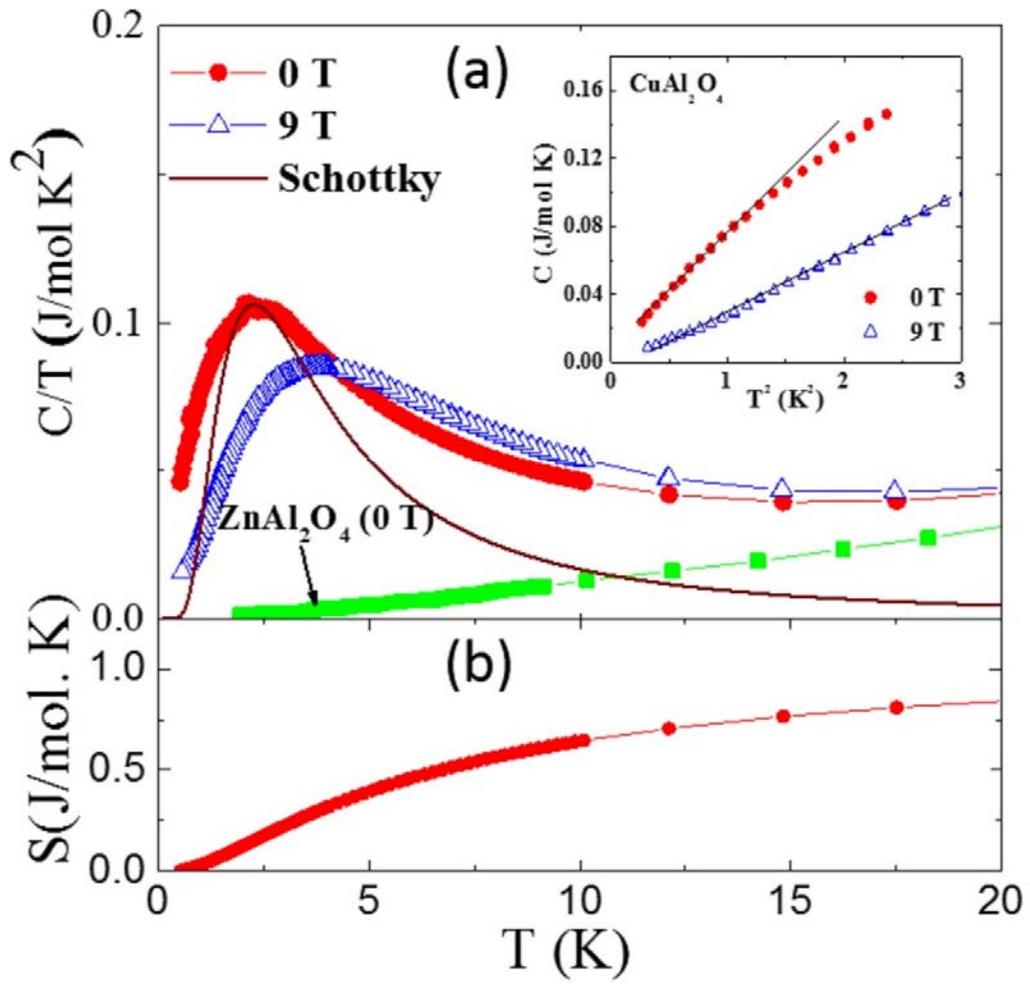

Figure 3

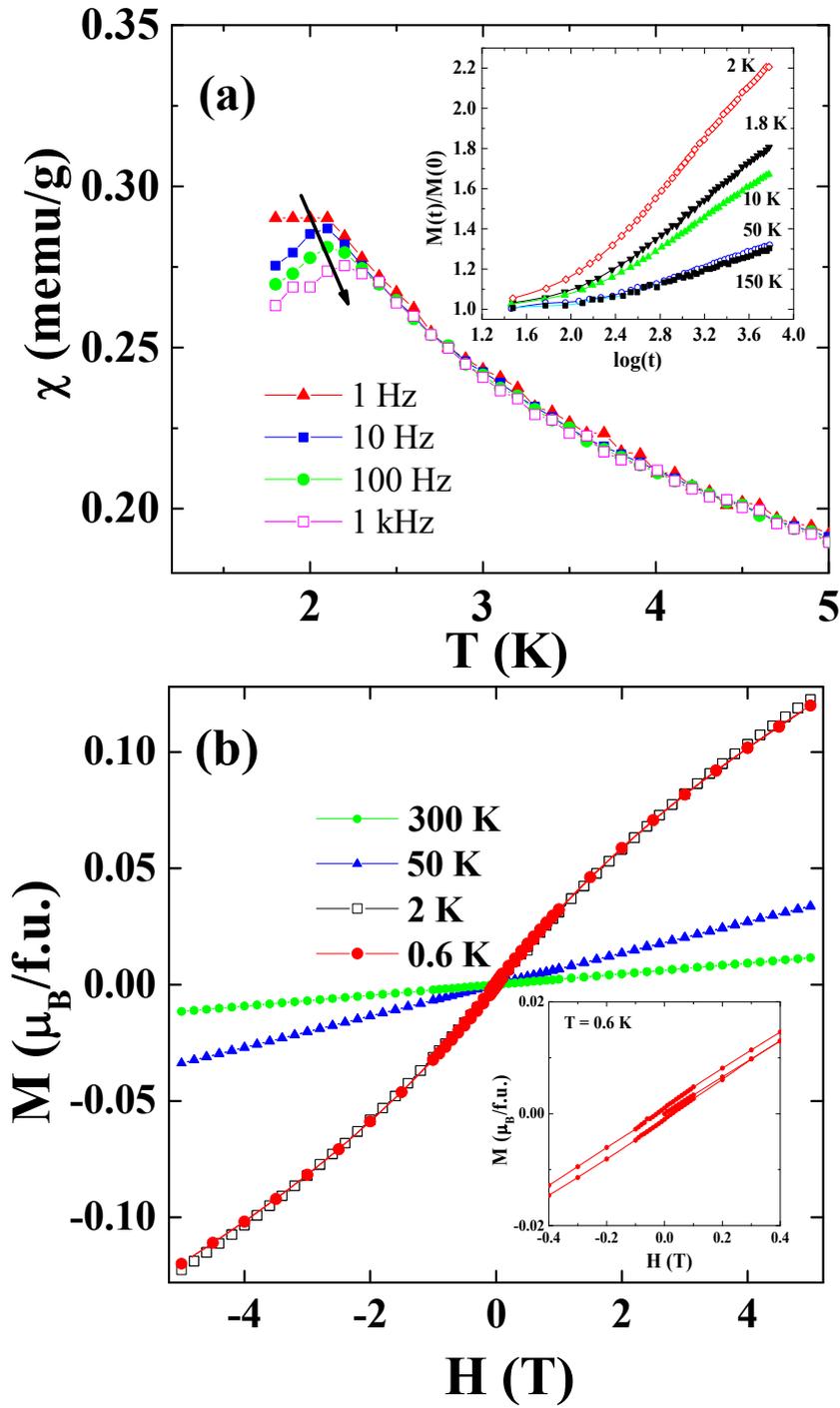

Figure 4

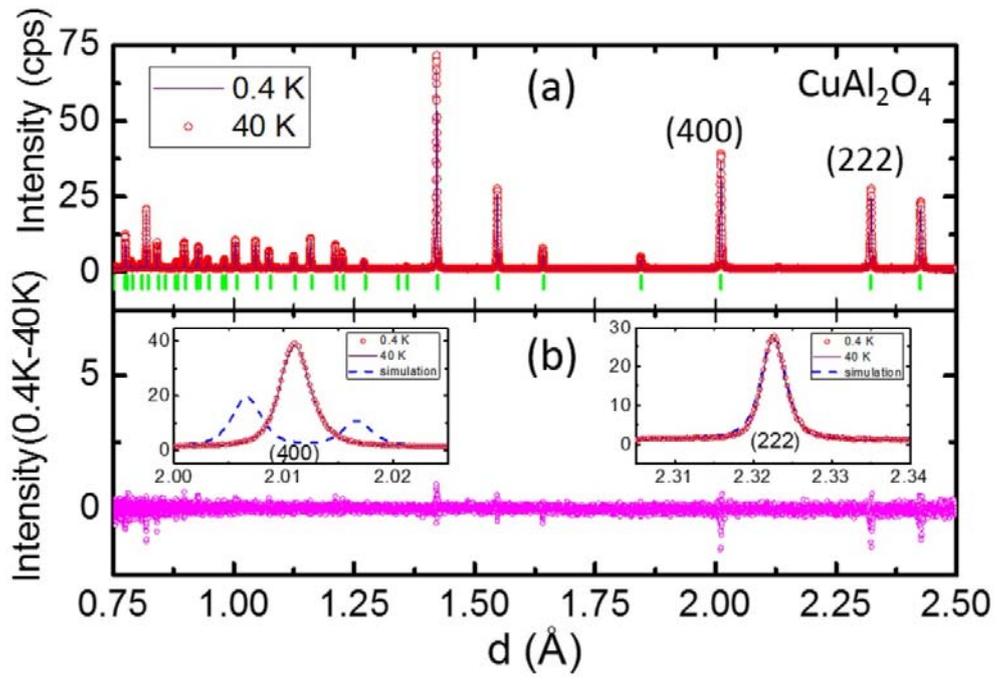

Table 1 Summary of the structural information. Bond distance at 40 K are as follows: Cu-Cu 3.494335(7); Cu-O 1.9104(6); Al-O 1.9278(6) Å. The position (0, 0, 0) are in tetrahedral sites and (5/8, 5/8, 5/8) are in octahedral site. In our final refinement, we achieved the following goodness of agreements: $R_p$=8.54, $R_{wp}$=9.62, $R_{exp}$=4.43, and $\chi^2$=4.71.

| 300 K | F d -3 m origin choice 1,  a = b = c = 8.07683(10) Å | | | | |
|---|---|---|---|---|---|
|  | x | y | z | B | Occ |
| Cu | 0 | 0 | 0 | 1.105(159) | 0.029 |
| Al | 0 | 0 | 0 | 1.105 (159) | 0.013 |
| Al | 0.625 | 0.625 | 0.625 | 0.505(155) | 0.070 |
| Cu | 0.625 | 0.625 | 0.625 | 0.505(155) | 0.013 |
| O | 0.38622(104) | 0.38622(104) | 0.38622(104) | 1.056(301) | 0.167 |

| 40 K | F d -3 m origin choice 1,  a = b = c = 8.06983(3) Å | | | | |
|---|---|---|---|---|---|
|  | x | y | z | B | Occ |
| Cu | 0 | 0 | 0 | 0.779(27) | 0.029 |
| Al | 0 | 0 | 0 | 0.779(27) | 0.013 |
| Al | 0.625 | 0.625 | 0.625 | 0.693(27) | 0.070 |
| Cu | 0.625 | 0.625 | 0.625 | 0.693(27) | 0.013 |
| O | 0.38645(5) | 0.38645(5) | 0.38465(5) | 0.131(18) | 0.167 |